\documentclass[pre,twocolumn,showpacs,floatfix]{revtex4}
\usepackage{graphics}
\usepackage{bm}
\usepackage{amsmath,amssymb}
\usepackage{cases}
\usepackage{amsfonts}
\begin{document}
\title{Phenomenological model for ordered onions under shear flow}
\author{Kenta Odagiri and Kazue Kudo}
\affiliation{Division of Advanced Sciences, Ochadai Academic Production,
Ochanomizu University, Tokyo 112-8610, Japan}

\begin{abstract}
We propose a phenomenological model for the multi-lamellar vesicles (onions)
formation induced by shear flow.
In a nonionic surfactant (C$_{12}$E$_4$) system, onion phases under a fixed
shear flow within a certain range show the order-disorder transition
accompanied with a size jump by changing temperature.
Our model can simulate ordered and disordered onion phases with different onion
sizes.
We show numerical results of the onion formation simulated by the model and
also discuss what factors in this system are critical to cause the transition
between these two different onion phases.
\end{abstract}

\pacs{05.65.+b, 82.70.Uv, 05.45.-a}

\maketitle

\section{Introduction}
\label{intro}

Surfactants in water form various kinds of assemblies such as lamellar,
micelles and vesicles. Multi-lamellar vesicles, which are also called onions,
are one of the interesting morphologies induced by shear flow.
The onion formation under shear flow has been, in fact, observed in experiments
\cite{Diat93,Diat95,Nettesheim,Fujii}.
Onions fill space with various sizes and forms of polyhedra.
Each onion consists of concentric lamellar membranes, which were observed by
electro-microscopy \cite{Gulik}.
The size of an onion depends on the shear rate and the characteristics of a
surfactant membrane.
Several theories have been proposed to estimate onion size.
Diat \textit{et al.} proposed the model in which applied shear stress is
balanced by the elastic stress given by the curvature energy of membrane
\cite{Diat93}.
On the other hand, van der Linden \textit{et al.} proposed the idea that
applied shear stress is balanced by the energy cost to deform an onion
\cite{Linden}.
These models have been examined by experiments especially concerning shear-rate
dependence.
The formation of onions, i.e. closed-packed or disordered structure, is another
problem to be investigated.
Panizza \textit{et al.} suggested that the disordered onion phase is made of
randomly oriented monodomains which consist of closed-packed onions
\cite{Panizza96}.

The characteristics of a surfactant membrane, e.g. elasticity, depend on
temperature as well as substances.
An interesting behavior related to the temperature dependence has been observed
in the shear-induced ordering of onions in a nonionic surfactant
(C$_{12}$E$_4$) system: Onions form a two-dimensional (2D) close-packed
honeycomb structure in a certain temperature interval and they are disordered
below and above the temperature interval \cite{Le,Suganuma}.
Moreover, in recent experiments, it has been revealed that the order-disorder
transition is accompanied by a size jump \cite{Suganuma}.
The size of onions is 5-6 times larger in the ordered phase than in the
disordered one.
To our knowledge, no theoretical model about the transition has been proposed,
though the order-disorder transition with a size jump had already been reported
in other systems \cite{Diat95,Sierro}.

In this paper, we propose a model to simulate ordered and disordered onion
phases with different onion sizes from the view point of pattern formation.
The model provides the means to demonstrate the formation of an onion phase in
real space.
Experiments by the small-angle light scattering (SALS) and small-angle X-ray
scattering (SAXS) give only images in reciprocal space.
It is difficult in those scattering experiments to see what structure onions
form in three dimensions.
Simulations in real space will give a clue to analyze the structure of the
onion phase which is observed experimentally.
In this paper, we focus on the onion formation induced by shear flow.
Our model illustrates the onion pattern which is stable under shear flow,
although the model is not suitable to discuss the lamellar-onion transition.

From the view point of pattern formation, 2D hexagonal structure such as the
ordered onion phase appears ubiquitously.
One of the most famous examples is nonequilibrium fluid dynamics known as
Rayleigh-B\'enard convection, of which pattern formation is well described by
the Swift-Hohenberg (SH) equation.
The SH equation is a well-known model equation for nonequilibrium pattern
formation phenomena \cite{Cross}, and it has a property of so-called potential
dynamics.
We derive a phenomenological model, which is similar to the SH equation, of the
onion formation by constructing the free energy of onion phases under shear
flow.

The rest of this paper is organized as follows. In Sec. \ref{model}, we
introduce a phenomenological model describing the onion formation induced by
shear flow.
In Sec. \ref{results}, we show numerical results of ordered and disordered
onion phases with different onion sizes. Section \ref{discussion} is devoted to
the discussion about how the onion size and the ordering structure of onions
are selected.
Finally, we conclude this paper in Sec. \ref{conclusion}.

\section{Model}
\label{model}

We here construct a phenomenological model for the shear-induced onion phase.
We mainly focus on the transition of the onion formation (size and ordering
structure) and do not consider the transition between lamellar and onion phases
in our model.
To simulate the onion formation under shear flow, we derive the free energy of
onion phases including the contribution from shear flow, so that we can realize
the stationary distribution of onions by minimizing the free energy.
 
\begin{figure}
\centering
\resizebox{1.0\columnwidth}{!}{
  \includegraphics{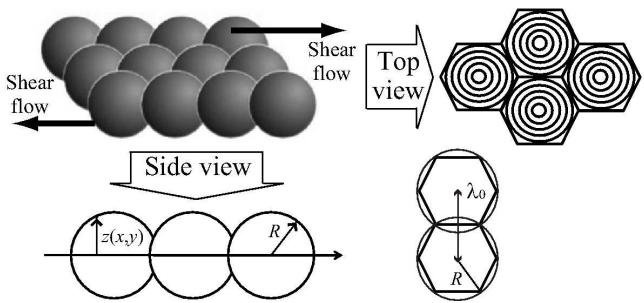}
}
\caption{Schematic representation of the shear-induced onion phase in our model.}
\label{fig:1}
\end{figure}

We first assume that onions, which have a certain identical size $R$, are
packed to form a 2D board.
The centers of onions are placed on the $x$-$y$ plane.
The spatial distribution of onions is expressed as the distribution of the
height of onion surface, $z(x,y)$ (see Fig. \ref{fig:1}).

\begin{figure}
\centering
\resizebox{0.5\columnwidth}{!}{
  \includegraphics{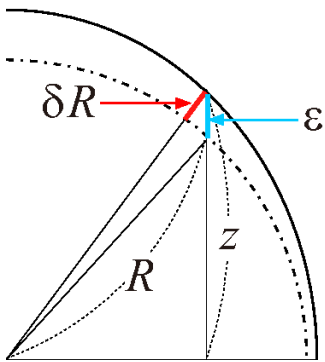}
}
\caption{(Color online) Schematic representation of a distorted vesicle
surface.
Solid and dot-dash curves denote the surface of a distorted vesicle and that of
a spherical one, respectively.}
\label{fig:2}
\end{figure}

Second, considering onions are closed-packed to form a 2D honeycomb structure
(Fig. \ref{fig:1}), we assume that the shape of an onion is a
slightly-distorted sphere (see Fig. \ref{fig:2}) because spherical onions
cannot fill space without deformations.
In fact, actual onions show a polyhedral structure in the experimental
observation \cite{Gulik}, and the distortion of the membranes is localized at
corners.
However, we assume the distortion of membranes is uniformly distributed over
slightly-distorted spherical vesicles for simplicity.
The validity of this assumption is discussed in the Sec. \ref{discussion}.

\subsection{Free energy for the onion phase under shear flow}
\label{sec:2-1}

On the above assumptions, we give the following Gibbs free energy $G$
including the contribution from shear flow.
\begin{align}
G=F_{\text{onion}} + F_{\text{surface}} - \sigma A.
\end{align}
The free energy $G$, which is a function of $z(x,y)$, consists of two parts:
(i) the contribution from the free energy of a deformed onion
($F_{\text{onion}}$) and (ii) the contribution from the stress applied to the
surface membrane of packed onions under shear flow
($F_{\text{surface}} - \sigma A$).
We derive these two parts of the free energy in what follows.

\subsubsection{Contribution from a deformed onion}
\label{sec:2-1-1}

The free energy of a deformed onion $F_{\text{onion}}$ consists of bending
energy $F_{\text{bend}}$ and interaction energy $F_{\text{int}}$ coming from
compression of the layers.
We here derive these two kinds of free energy as follows.

We first derive $F_{\text{bend}}$ in a similar manner to that of Ref.
\cite{Ramos}.
The bending free energy per area $F/A$ for a bilayer membrane is represented as
the following function of the mean and Gaussian curvatures, $H$ and $\bar{H}$,
respectively.
\begin{align}
\frac{F}{A}=2\kappa H^2 + \bar{\kappa}\bar{H} + \frac{1}{4} c_1 H^4
+ \frac{1}{4} c_2 \bar{H}^2 + 2 c_3 H^2 \bar{H},
\end{align}
where $\kappa$ and $\bar{\kappa}$ are the conventional bending and Gaussian
curvature moduli, respectively, $c_1, c_2$ and $c_3$ are nonlinear moduli.
We here consider a vesicle which has a spherical shape, which indicates that
$H^2$ and $\bar{H}$ are equivalent. Thus, the free energy,
$F_{\text{shell}}(r)$, of a spherical shell with radius $r$ is given by
\begin{align}
F_{\text{shell}}(r) = 4\pi \tilde{\kappa} H^2 r^2 + \pi \tilde{c} H^4 r^2,
\label{eq:f_shell}
\end{align}
where $\tilde{\kappa} \equiv 2\kappa + \bar{\kappa}$ and
$\tilde{c} \equiv c_1 + c_2 + 8c_3$.

We next consider distortion of the spherical shell due to packed-onion
formation.
We assume that the radius of curvature changes from an ideal radius $r$ to that
for a distorted spherical shell $r+\delta r$ (Fig. \ref{fig:2}).
Since the degree of distortion is quite smaller than the ideal radius
($r \gg \delta r$), we expand the mean curvature $H$ in $\alpha=\delta r/r$ and
keep terms up to the third order.
\begin{align}
H=\frac{1}{r + \delta r} \simeq \frac{1}{r}(1 - \alpha +\alpha^2 - \alpha^3).
\end{align}
Similarly, we have $H^2$ and $H^4$ as follows.
\begin{align}
H^2 &= \frac{1}{r^2}(1 - 2\alpha + 3\alpha^2 - 4\alpha^3), \\
H^4 &= \frac{1}{r^4}(1 - 4\alpha + 10\alpha^2 - 20\alpha^3).
\end{align}
Substituting these expressions into Eq. (\ref{eq:f_shell}), we have
\begin{align}
F_{\text{shell}}(r) &= 4\pi \tilde{\kappa} (1 - 2\alpha + 3\alpha^2 - 4\alpha^3) \notag \\
&\quad + \frac{\pi \tilde{c}}{r^2} (1 - 4\alpha + 10\alpha^2 - 20\alpha^3).
\end{align}
Thus, the free energy $f_{\text{onion}}(R)$ (per unit volume of an onion with
radius $R$ made of $n$ vesicles with layer spacing $d$) is given by
\begin{align}
f_{\text{onion}}(R=nd) &= \sum_{j=1}^{n} \frac{F_{\text{shell}}(r=jd)}{\frac{4}{3}\pi R^3}  \notag \\
&= \frac{3\tilde{\kappa}}{dR^2}(1 - 2\bar{\alpha}
+ 3\bar{\alpha}^2 - 4\bar{\alpha}^3) \notag \\
&\quad + \frac{3\tilde{c}}{4d^2 R^3}\left( 1-\frac{d}{R}\right)(1 - 4\bar{\alpha}
+ 10\bar{\alpha}^2 - 20\bar{\alpha}^3),
\end{align}
where $\bar{\alpha}=\delta R/R$ is the ratio of distortion on the outermost
shell of an onion.
Let us consider the distortion of the onion surface.
As illustrated in Fig. \ref{fig:2}, the difference between the height of the
distorted outermost shell and that of the spherical one is taken to be
$\epsilon$, which leads the following relation:
$R^2-(z-\epsilon)^2=(R+\delta R)^2-z^2$.
Under the conditions of slight distortion ($R\gg \delta R$ and
$z \gg \epsilon$), we obtain the following relation
\begin{align}
\bar{\alpha} = \delta R / R = (\epsilon/R^2)z \equiv \tilde{\alpha}z.
\end{align}

By using the above relation, we finally have $F_{\text{bend}}$ as a function
of $z(x,y)$,
\begin{align}
F_{\text{bend}} &= \int dxdy \Bigl[ f_{\text{onion}}(R) \cdot z\Bigr] \notag \\
&= \int dxdy \Bigl[ \frac{3\tilde{\kappa}}{dR^2}(z - 2\tilde{\alpha}z^2
+3\tilde{\alpha}^2z^3 - 4\tilde{\alpha}^3z^4) \notag \\
&\quad + \frac{3\tilde{c}}{4d^2 R^3}\left( 1 - \frac{d}{R}\right)(z - 4\tilde{\alpha}z^2
+ 10\tilde{\alpha}^2z^3 - 20\tilde{\alpha}^3z^4) \Bigr] .
\end{align}

We next derive the interaction energy $F_{\text{int}}$ related to compression.
The lammellar interaction between layers is given by $\frac{1}{2}B_0d^2$, where
$B_0$ and $d$ are the compression modulus and layer spacing, respectively.
Applying the interaction to layers inside a onion with radius $R$, the
interaction between vesicle layers depends the height of the onion surface
$z(x,y)$ because effective layer spacing is given by $\frac{z}{R}d$.
We thus have $F_{\text{int}}$ as a function of $z(x,y)$,
\begin{align}
F_{\text{int}} = \int dxdy \left[ \frac{n}{2}B_0 \left( \frac{z}{R}d \right)^2 \right]
= \int dxdy \left( \frac{B_0d}{2R} z^2 \right) ,
\end{align}
where we have used $n=R/d$.

From the above derivation, we finally obtain the free energy of a deformed
onion $F_{\text{onion}}$
\begin{align}
F_{\text{onion}} &= F_{\text{bend}}+F_{\text{int}} \notag \\
&= \int dxdy \Bigl[ \frac{3\tilde{\kappa}}{4d^2R^3} \bigl(
-(4+20\lambda)\tilde{\alpha}^3 z^4 +(3+10\lambda)\tilde{\alpha}^2 z^3 \notag \\
&\quad -(2+4\lambda-\frac{\bar{B_0}}{2})\tilde{\alpha}z^2 +(1+\lambda)z \bigr) \Bigr] ,
\end{align}
where $\lambda=\frac{(R-d)\tilde{c}}{4dR^2\tilde{\kappa}}$ and
$\bar{B_0}=\frac{4B_0d^3R^2}{3\tilde{\alpha}}$.

\subsubsection{Contribution from the stress applied to the surface membrane
under shear flow}
\label{sec:2-2-2}
We next consider the free energy from the stress applied to the surface
membrane of packed onions under shear flow.
Suppose that onion surfaces are covered by a single membrane as if the membrane
is a part of each onion surface, and also that the membrane is subjected to an
effective stress that comes from surface tension and shear stress.
From these assumptions, the free energy contribution from the effective stress
is written as
\begin{align}
F_{\text{surface}} - \sigma A = \int dA (2\kappa H^2 - \sigma),
\label{eq:stress}
\end{align}
where $\sigma$ and $A$ denote the effective stress and the total area of the
membrane, respectively.
The first term is the bending free energy of the onion surface as a single
membrane and the second term is the effective stress applied to the membrane
under shear flow.
We assume the effective stress $\sigma$ includes contributions not only from
effective surface tension but also from shear stress.
Although the shear can affect vesicle membranes inside onions, we impose the
effect from shear stress on only the surface membrane for simplicity.

Using $\bm{s}=\mathbf{\nabla} z$, we have the following expressions \cite{Peliti}
\begin{align}
dA &= \sqrt{1+\bm{s}^2}dxdy, \\
H &= \frac{\mathbf{\nabla} \cdot \bm{s}}{\sqrt{1+\bm{s}^2}}.
\end{align}
Substituting these expressions into Eq. (\ref{eq:stress}) and assuming
$\bm{s}^2 \ll 1$, we finally obtain
\begin{align}
F_{\text{surface}} - \sigma A = \int dxdy\Bigl[ 2\kappa (\nabla^2 z)^2
- \frac{\sigma}{2} (\nabla z)^2 - \sigma \Bigr] .
\end{align}

\subsection{Distribution of the onion surface $z(x,y)$}
\label{sec:2-2}

To simulate the stationary distribution of the onion surface $z(x,y)$, we use
the following Ginzburg-Landau (GL) equation \cite{Onuki}, in which time
evolution is given by the variation of the free energy $G$,
\begin{align}
\frac{\partial z}{\partial t}=-L\frac{\delta G}{\delta z}
= -\tilde{f}_{\text{onion}}(z) + \sigma\nabla^2 z - 4\kappa\nabla^4 z,
\label{eq:z_evo}
\end{align}
where
\begin{align}
\tilde{f}_{\text{onion}}(z)
&= \frac{3\tilde{\kappa}}{4d^2R^3} \Bigl[ -(16+80\lambda)\tilde{\alpha}^3 z^3
+(9+30\lambda)\tilde{\alpha}^2 z^2 \notag \\
&\quad -(4+8\lambda-\bar{B_0})\tilde{\alpha}z +(1+\lambda) \Bigr].
\label{eq:f_onion}
\end{align}
Here, the phenomenological coefficient $L$ has the dimension of time over mass
and we fix $L=1$ in the following. By solving the above equation, we finally
obtain the stationary distribution of the onion surface $z(x,y)$.
Since the stationary distribution is realized by minimizing the free energy $G$
of onion phases under shear flow, we can simulate the onion pattern formation
under shear flow by solving Eq. (\ref{eq:z_evo}).

It should be noted that Eq. (\ref{eq:z_evo}) has essentially the same form as
the SH equation.
In other words, the equation we have derived is in the same class as the SH
equation, although the mechanism of onion pattern formation is different from
that of the system in which the SH equation was originally derived.

\section{Results}
\label{results}

We here show numerical results obtained from our model to demonstrate the
change between a small-disordered onion phase and a large-ordered one. Table
\ref{tab:1} lists the system parameters of the model. The system size is
$200 \times 200$ (lattice) with the periodic boundary condition. To integrate
the model equation, we have used the Crank-Nicolson scheme. Since we are mainly
interested in the packed-onion ordering, we focus only on the stationary
distribution of $z(x,y)$ after a sufficiently long time.
In our numerical simulations, $\kappa$ is fixed ($\kappa = 10.0$), because it
is considered to be almost constant under experimental conditions for onion
phases \cite{Suganuma2}. $B_0$ and $R$ is set by using following relations:
$B_0=\sigma^2 d / \kappa$ and $R=4\pi\sqrt{2\kappa / 3\sigma}$. The validity of
these relations is discussed in the next section. Two different sets of the
other parameters (cases A and B) are listed in Table \ref{tab:2}.

\begin{table}
\caption{System parameters}
\label{tab:1}
\begin{center}
\begin{tabular}{clcl}
\hline\hline
$\sigma$ & effective stress & ~~$B_0$ & compression modulus \\
$\kappa$ & bending modulus & ~~$R$ & radius of an onion \\
$\bar{\kappa}$ & saddle-splay modulus & ~~$d$ & layer spacing \\
$\tilde{c}$ & nonlinear modulus & ~~$\epsilon$ & shift of $z$ caused \\
 & & ~~ & by distortion \\
\hline\hline
\end{tabular}
\end{center}
\end{table}

\begin{table}
\caption{System parameters chosen for large (case A) and small (case B) onion
phases.}
\label{tab:2}
\begin{center}
\begin{tabular}{lcccccccccc}
\hline\hline
&& $\sigma$ && $\bar{\kappa}$ && $\tilde{c}$ && $d$ && $\epsilon$ \\
case A && $20.0$  && $-13.0$ && $-3.9$ && $0.014$ && $1.20$ \\
case B && $200.0$ && $-1.0$  && $-3.0$ && $0.013$ && $0.35$ \\
\hline\hline
\end{tabular}
\end{center}
\end{table}

\begin{figure}
\centering
\resizebox{0.95\columnwidth}{!}{
  \includegraphics{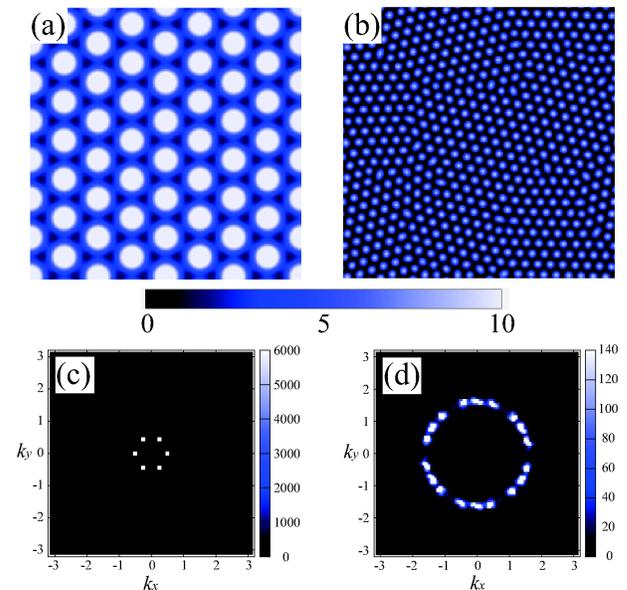}
}
\caption{(Upper panels) Spatial distributions of $z(x,y)$ for (a) case A, and
(b) case B. The brighter the color, the larger the value of $z$. (Lower panels)
Spatial power spectra for (c) case A and (d) case B.}
\label{fig:3}
\end{figure}

Figure \ref{fig:3} shows the distributions of $z(x,y)$ and spatial power
spectra for case A and case B. For case A, the size of onions is large
($R\simeq 7.3$) and onions show a honeycomb structure (panel (a)). By contrast,
for case B, the size of onions is quite small ($R\simeq 2.3$), although the
distribution of onions looks an almost perfect honeycomb structure (panel (b)).
However, spatial power spectra for these cases are totally different. Panel (c)
shows a perfect six-fold symmetry for case A, although panel (d) shows an
isotropic circular spectrum for case B. This isotropic ring indicates that the
onion phase consists of some small domains of packed onions (i.e. onion
texture, \cite{Panizza96,Courbin}): Each small domain has the perfect six-fold
symmetry but has different directions. These random directions of small domains
result in the isotropic ring spectrum.

Furthermore, the time duration before the distribution of $z(x,y)$ falls into a
stationary state is totally different for cases A and B. While the ordered
onion phase gradually appears in case A, some small domains appear very quickly
in case B. The difference can be explained in terms of the linear-growth rate,
which is estimated from the linear stability analysis. Since the linear-growth
rate is a very small positive value in case A, the stationary pattern can reach
the potential minimum, where the ordered honeycomb structure is stable. On the
other hand, in case B, because the linear-growth rate is a relatively large
positive value, the stationary pattern quickly reaches a metastable state,
where some randomly-directed monodomains coexist, and stays there.

We have also checked the robustness of these results by simulations in a larger
system size. These results indicate that our model is able to reproduce the
change between the large-ordered onion phase and the small-disordered one by
changing the system parameters. We next discuss the validity of changing these
parameters.

\section{Discussion}
\label{discussion}

We here discuss what factors in our model are essential to induce the change
between the large-ordered onion phase and the small-disordered one, focusing on
tha validity and temperature dependence of the parameters in our model.

\subsection{Onion size}
At first, we focus on how to determine onion size $R$ in our model. As
mentioned above, Eq. (\ref{eq:z_evo}) has a similar mathematical structure to
that of the SH equation, which is a model equation to simulate periodic domain
patterns. The characteristic length $\lambda_0$ of the domain pattern produced
by the SH equation can be estimated effectively from the linear stability
analysis. In the case of Eq. (\ref{eq:z_evo}),
$\lambda_0 =2\pi \sqrt{8\kappa / \sigma}$. $\lambda_0$ represents the distance
between the centers of neighboring onions and it also has a relation
$\lambda_0=\sqrt{3}R$ because of the geometrical structure of packed onions
(see Fig. \ref{fig:1}). Therefore, the estimated onion size is given by
\begin{align}
R = \frac{\lambda_0}{\sqrt{3}}=4\pi\sqrt{\frac{2\kappa}{3\sigma}}.
\label{eq:R}
\end{align}
This equation indicates that the onion size in our model is determined by the
ratio of $\kappa$ and $\sigma$. We set $R$ in our simulations by using Eq.
(\ref{eq:R}).

Actually, only $\sigma$ is the essential parameter to determine the
onion size, because $\kappa$ is fixed in our simulations. Under a fixed shear
rate, it is assumed that the change of $\sigma$ is caused by the change of the
surface tension. According to experimental studies \cite{Panizza96,Leng}, the
elastic modulus $G'$ measured in linear viscoelastic experiments is associated
with the surface tension $\sigma_{\text{s}}$ and onion size $R$:
$G' \sim \sigma_{\text{s}} / R$. Assuming $\sigma_s \simeq \sigma$ and
substituting Eq. (\ref{eq:R}) into the above relation, we have
$G' \sim \sigma_s^{\frac{3}{2}}$. In the experiment of the
$\text{C}_{12}\text{E}_4$ system, on the other hand, $G'$ decreases as the
radius of onions increases \cite{Suganuma2}. These experimental results
indicate that $\sigma_s$ in the large-ordered onion phase is smaller than that
in the small-disordered one. This property of the surface tension $\sigma_s$
corresponds qualitatively to that of the parameter $\sigma$ in our simulations
($\sigma=20.0$ and $200.0$ in cases A and B, respectively).

\begin{figure}
\centering
\resizebox{0.85\columnwidth}{!}{
  \includegraphics{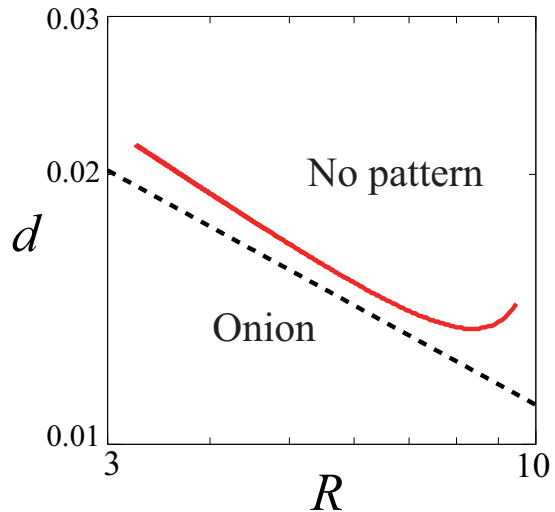}
}
\caption{Phase diagram of onion phase as a function of onion size $R$ and
layer spacing $d$.
The solid curve indicates the boundary between the onion phase ("Onion") and a
phase where no pattern appears ("No pattern"), which is calculated from the
linear stability analysis.
The slope of the dashed line is $-\frac{1}{2}$.}
\label{fig:4}
\end{figure}

We further discuss the change in onion size as a function of shear rate
$\dot{\gamma}$ and layer spacing $d$.
Diat \textit{et al.} estimated onion size $R$ by considering a balance between
the elastic stress and the viscous stress as follows \cite{Diat93}.
\begin{align}
R \simeq \sqrt{\frac{4\pi \bar{\kappa}}{\eta d \dot{\gamma}}},
\end{align}
where $\eta$ is the solvent viscosity.
For a fixed $\eta$ (i.e. Newtonian fluid), this relation implies
$R \sim (d \dot{\gamma})^{-\frac{1}{2}}$.
This estimation accords with $\dot{\gamma}^{-\frac{1}{2}}$ dependence
experimentally observed by themselves.
Assuming the effective stress $\sigma$ in our model is estimated as
$\sigma \sim \eta \dot{\gamma}$ and substituting it into Eq. (\ref{eq:R}), we
also have $R \sim \dot{\gamma}^{-\frac{1}{2}}$ for a fixed $\eta$.
We further estimate the relation between onion size $R$ and layer spacing $d$
from the linear stability analysis of Eq. (\ref{eq:z_evo}).
Figure \ref{fig:4} shows a phase diagram of onion phase as a function of $R$
and $d$.
The boundary of onion phase, which is calculated from the linear stability
analysis, has a slope $-\frac{1}{2}$.
It indicates the relation $R \sim d^{-\frac{1}{2}}$, because the ordered onion
phase appear near the boundary, that is, the region where the linear-growth
rate is very small.

Actually, the order-disorder transition with a size jump is related to the
compression of the onion layers.
It causes a discontinuous change of the viscosity \cite{CourbinPRL}.
The experimental result of Ref. \cite{CourbinPRL} also shows the shear
thinning, that is, the viscosity $\eta$ depends on the shear rate
$\dot{\gamma}$.
However, the effective stress $\sigma$ in our model includes not only the shear
stress but also $\eta$, $\dot{\gamma}$, and $d$.
Thus, we cannot discuss the compression or shear thinning in our model.

\subsection{Degree of the distortion}
We next consider the validity of the distortion of vesicle membranes due to
packing of onions. Although we assume that membranes of packed onions are
slightly distorted from a sphere, actual onions have a polyhedral structure
\cite{Gulik}. 
We clarify the validity of
the assumption by estimating the degree of the distortion $\epsilon / R$.
Inside an onion formed by polyhedral vesicles, the effective surface tension
$\sigma_\text{eff}$ and the penetration length $\xi$, which corresponds to the
distortion of vesicle membranes, can be estimated as
$\sigma_\text{eff} \simeq \sqrt{KB}$ and $\xi \sim \sqrt{K/B}$, respectively,
where $K$ is the bending modulus $K=\kappa / d$ and $B$ is the compression
modulus \cite{Panizza96,Linden2}. 
We here assume that the effective surface
tension corresponds approximately to the stress $\sigma$ in our model, namely
$\sigma_\text{eff} \simeq \sigma$. We thus have $\xi\sim \kappa /\sigma d$.
In addition, we can estimate the compression modulus $B_0$ in our simulation by
using the above estimation of $\sigma_{\text{eff}}$, and consequently, we set
$B_0=\sigma^2 d / \kappa$.

On the other hand, the onion size $R$ can be estimated to be
$R\sim \sqrt{\kappa / \sigma}$ as mentioned above. We thus obtain the following
relation: $\xi / R \sim R$. This indicates that the larger the onion size $R$
the larger the distortion $\epsilon$, if one assumes $\xi \sim \epsilon$. The
above relation is also equivalent to the relation $\xi / R^2 =\text{constant}$.
This corresponds to our assumption
$\tilde{\alpha} = \epsilon / R^2 = \text{constant}$. Therefore, the assumption
for the distortion of vesicle membranes due to packing of onions is consistent
with other assumptions. In addition, we set $\epsilon$ in our simulations to
satisfy the above relation that the larger onion has larger distortion.

\subsection{Elastic properties of vesicle membranes in packed onions}
Elastic properties of vesicle membranes in packed onions, such as $\kappa$,
$\bar{\kappa}$, and $\tilde{c}$, should also be considered. We first focus on
the signs of coefficients of the bending free energy such as $\kappa$ and
$\bar{\kappa}$. In general, the saddle-splay modulus $\bar{\kappa}$ is a
negative value unless the sponge ($L_3$) phase is stable and spherical vesicles
are stabilized for $\tilde{\kappa}=2\kappa +\bar{\kappa}<0$. However, the
effects of shear flow are not taken into account in this condition. Thus, it is
not necessary in our model to satisfy the condition. We have actually checked
simulations for different sets of parameters $\kappa$, $\bar{\kappa}$, and
$\tilde{c}$ and consequently found that the onion phase can be obtained only
when $\bar{\kappa}<0$ and $\tilde{\kappa}>0$. This result suggests that the
shear stress stabilizes unstable spherical vesicles for $\tilde{\kappa}>0$.

We further refer two important nonlinear effects of the bending free energy to
simulate onion phases in our model: nonlinear modulus in the bending energy and
the distortion of membranes due to packing of onions. The usual expression of
the Helfrich curvature free energy does not include the nonlinear coefficient
$\tilde{c}$. However, $\tilde{c}$ should be considered in an onion phase,
because high curvatures are expected near the core of an onion \cite{Ramos}.
Also in our model, $\tilde{c}$ is one of the essential parameters to simulate
onion phases, because no onion phase is generated from our model if $\tilde{c}$
is taken to be $0$. With regard to the distortion of membranes, it is also
necessary to consider the distortion of vesicle membranes in our model to
simulate onion phases. The third-order expansion of the curvature on a
distorted vesicle is necessary to generate the ordered onion formation, because
the non-linear terms of $z$ in Eq. (\ref{eq:f_onion}) play a similar role of a
double-well potential.

\subsection{Temperature dependence of the parameters}
We further consider the relation between the temperature dependence of the system
parameters in our simulations and that of the physical quantities in
experimental studies. Since $\sigma$ is the critical parameter to determine the
onion size in our simulations, it should have a similar temperature dependence
to that of the onion size. In other words, $\sigma$ should be minimized in a
certain temperature interval and have larger values above and below the
interval.

On the other hand, Kosaka \textit{et al} \cite{Kosaka} reported that
layer spacing $d$ can change around the lamellar-onion transition, and also
discussed temperature dependence of $\bar{\kappa}$. They measured $d$ at
different concentrations of the surfactant and showed that the ratio
$\delta / d$ of $d$ and the thickness of bilayers $\delta$ has a temperature
dependence. Although they focused on the lamellar-onion transition in their
experiments, the temperature dependence of $\delta / d$ may be one of the
important factors to cause the order-disorder transition in onion phases.
We therefore change the value of $\bar{\kappa}$ and $d$ in our simulations,
assuming the temperature dependence of these parameters. Actually, they are not
essential parameters to determine the onion size, but important parameters to
determine the linear-growth rate, that is, determine whether onion phases
appear or not.




\section{Conclusion}
\label{conclusion}

We have proposed the phenomenological model for the onion pattern formation
induced by shear flow. We have derived the free energy of an onion phase by
considering contributions from deformed packed onions and from the stress
applied to the surface membrane on packed onions under shear flow. Our model
includes two important nonlinear effects, that is, nonlinear modulus
$\tilde{c}$ in the bending energy and the distortion of membranes due to
packing of onions. These nonlinearity are crucial in simulations to reproduce
onion phases in our model. Then, we have demonstrated that the two different
onion phases, large-ordered one and small-disordered one, can be simulated by
adjusting the system parameters in our model. The ordered onion phase has a
clean honeycomb structure made of packed onions. The disordered onion phase
consists of small domains of ordered packed onions. Each domain has the
six-fold symmetry, and domains are laid in random directions. The size of
packed onions is determined by the ratio of the bending modulus $\kappa$ and
the effective stress $\sigma$. Thus, $\sigma$ is the main control parameter
under the fixed-$\kappa$ condition, though other parameters, such as
$\bar{\kappa}$ and $d$, are also important for emergence of onion phases. In
fact, the system parameters in our model, such as $\sigma$, are difficult to
compare with the physical quantities which can be measured in experiments. In
that sense, our model can be improved, nevertheless, it is useful as a simple
model for the shear-induced onion formation.

\begin{acknowledgments}
The authors would like to thank Y. Suganuma, T. Kato and J. Fukuda for useful
comments and discussion.
\end{acknowledgments}


\end{document}